\documentclass[fleqn,12pt]{article}
\usepackage{amsmath,amssymb,epsfig,times,graphics}
\usepackage{amssymb}
\usepackage{amsmath}
\usepackage{CJK}
\usepackage{graphics}
\usepackage{graphicx}
\usepackage{dcolumn}
\usepackage{bm}
\usepackage{subfigure}
\usepackage{lineno}
\usepackage{multirow}
\usepackage{amsthm}
\begin{document}

\title{\bf A Bayesian statistical analysis of stochastic phenotypic plasticity model of cancer cells} \date{}
\maketitle

\author{Da Zhou$^{1}$, Shanjun Mao$^{2}$, Jing Cheng$^{3}$, Kaiyi Chen$^{1}$, Xiaofang Cao$^{1}$, Jie Hu$^{1,*}$}

\begin{enumerate}
  \item School of Mathematical Sciences, Xiamen University, Xiamen 361005, P.R. China
    \\ ($*$Corresponding Author, hujiechelsea@xmu.edu.cn))
  \item Department of Statistics,
The Chinese University of Hong Kong, Shatin, N.T., Hong Kong, PR China
  \item School of Statistics, Huaqiao University,
Xiamen 361005, P.R. China
\end{enumerate}

\section*{Abstract}
The phenotypic plasticity of cancer cells has received special attention
in recent years. Even though related models have been widely studied
in terms of mathematical properties, a thorough statistical analysis on
parameter estimation and model selection is still very lacking.
In this study, we present a Bayesian approach on the relative frequencies
of cancer stem cells (CSCs). Both Gibbs sampling and Metropolis-Hastings (MH) algorithm
are used to perform point and interval estimations of cell-state transition rates
between CSCs and non-CSCs. Extensive simulations
demonstrate the validity of our model and algorithm. By applying this method to a
published data on SW620 colon cancer cell line, the model selection
favors the phenotypic plasticity model, relative to
conventional hierarchical model of cancer cells. Moreover,
it is found that the initial state of CSCs after cell sorting significantly influences
the occurrence of phenotypic plasticity.

\section{Introduction}
\label{}

The hypothesis of cancer stem cell theory \cite{reya2001stem,jordan2006cancer} postulates a hierarchical organization of cancer cells.
A small number of tumorigenic cancer cells, also termed cancer stem cells (CSCs), reside at the apex
of this cellular hierarchy \cite{dalerba2007cancer}. CSCs are capable of self-renewal and generating more
differentiated cancer cells with lower tumorigenic potential. However, growing researches
have extended the CSC model by proposing a \emph{phenotypic plasticity} paradigm in which reversible
transitions could happen between CSCs and non-CSCs \cite{marjanovic2013cell}.
That is, not only can CSCs give rise to non-CSCs, but a fraction of non-CSCs can reacquire CSC-like
characteristics. This \emph{de-differentiation} from non-CSCs to CSCs has been reported in
quite a few types of cancer, such as breast cancer \cite{meyer2009dynamic,gupta2011stochastic,chaffer2013poised},
melanoma \cite{quintana2010phenotypic}, colon cancer \cite{yang2012dynamic},
and glioblastoma multiforme \cite{fessler2015endothelial}.

Very recently special attention has been paid to reasonable mathematical models for quantifying the process of phenotypic plasticity. In particular, it was found that the phenotypic plasticity plays an important role in the stability of the quantitative models
\cite{gupta2011stochastic,dos2013possible,dos2013noise,wang2014dynamics,
zhou2014multi,zhou2014nonequilibrium,niu2015phenotypic}.
That is, the phenotypic plasticity greatly contributes to
stabilizing the phenotypic mixture of cancer cells, thereby effectively maintaining the heterogeneity of cancer cell populations. Some other researches laid emphasis on the role of the phenotypic plasticity
in transient dynamics. It was shown that an interesting overshoot phenomenon of CSCs observed in experiment can be well explained by de-differentiation from non-CSCs to CSCs
\cite{sellerio2015overshoot,chen2016overshoot}. Besides, Leder et al studied mathematical models of
pdgf-driven glioblastoma and revealed that the effectiveness of radiotherapy
is quite sensitive to the capability of de-differentiation from differentiated sensitive
cells to stem-like resistant cells \cite{leder2014mathematical};
Jilkine and Gutenkunst studied the effect of de-differentiation on time to mutation acquisition
in cancers \cite{jilkine2014effect};
Chen et al studied transition model between endocrine therapy responsive and resistant states in breast cancer by
Landscape Theory \cite{chen2014mathematical};
Dhawan et al showed with mathematical modeling that exposure to hypoxia enhanced
the plasticity and heterogeneity of cancer cell populations \cite{dhawan2016mathematical};
Tonekaboi et al investigated how cellular plasticity behaves differently in small and large cancer cell populations
\cite{tonekaboni2017mathematical}.
A recent review by Jolly et al \cite{jolly2017epithelial} focused on quantitative models
of Epithelial-mesenchymal plasticity in cancer.

Even though the phenotypic plasticity has been extensively studied in terms of
mathematical properties, the statistical analysis on parameter estimation and
model selection is still very lacking. Actually, one of the crucial tasks in
the research of phenotypic plasticity is to estimate the
transition rates between different cell types. As a pioneering work, Gupta et al \cite{gupta2011stochastic} established a discrete-time Markov state transition model and estimated the transition probabilities between different cell states by fitting the model
to their FACS (Fluorescence-activated cell sorting) data on SUM159 and SUM149 breast cancer cell lines.
Besides, continuous-time ordinary differential equations (ODEs) models were also developed
\cite{wang2014dynamics,zhou2014multi}, based on which de-differentiation rates
were estimated by fitting to SW620 colon cancer cell line.
However, the above mentioned works can only provide point estimations
to the interested parameters, but not interval estimations.
Comparatively,
interval estimation is much more informative and frequently-used than point estimation in practice.
For doing interval estimation, statistical modeling rather than deterministic modeling should be applied.
Moreover, note that it is still questionable if
de-differentiation is a crucial improvement to the cellular hierarchy of cancer cells or just a minor extension to it, model selection can be used to validate the paradigm of phenotypic plasticity in terms of statistical significance. Therefore, a thorough statistical analysis is
of great value for further quantifying the biological process of phenotypic plasticity and exploring its biological significance.

In this research, a statistical framework is presented to analyze a two-phenotypic model of cancer cells. In this model, each cancer cell is either CSC phenotypic state or non-CSC phenotypic state. Both types of cells can divide symmetrically or asymmetrically with certain probabilities.
A Bayesian approach \cite{Hoff2009A} is developed to fit this model to experimental data on relative frequencies of CSCs. MCMC methods (such as Gibbs sampling \cite{Geman1987Stochastic} and MH algorithm \cite{Metropolis1953Equation,Hastings1970Monte}) are used to perform statistical inference with Multivariate Potential Scale Reduction Factor (MPSRF) \cite{gelman1992inference,Brooks1998general} checking the convergence of MCMC chains.
Our simulation results demonstrate the precision and accuracy of our algorithm by both point estimation and interval estimation.
By applying our approach to a published data on SW620 colon cancer cell line \cite{yang2012dynamic}, we also perform model selection via deviance information criterion (DIC, \cite{Gelman2003Bayesian}). Our result shows that the phenotypic plasticity model with de-differentiation has superior quality relative to the hierarchical model without de-differentiation. Furthermore, an interesting frequency-dependent phenomenon is presented, i.e. the estimated values of the model parameters depend on the initial relative frequencies of different cell states. This suggests that the process of phenotypic plasticity could be relevant to the heterogeneity level of cancer cell populations.

The paper is organized as follows. The model assumptions and Bayesian framework are presented in Section 2. Main results including simulations and real data analysis are
shown in Section 3. Conclusions are presented in Section 4.

\section{Methods}

\subsection{Model assumptions}\label{model}

In this section we describe the model assumptions. Note that the salient feature of the phenotypic plasticity model is the reversibility between CSCs and non-CSCs, i.e., not only can CSCs differentiate into non-CSCs, but non-CSCs are also capable of de-differentiating into CSCs. Consider a population of cancer cells comprising two phenotypes: CSC represents cancer stem cell phenotypic state,
non-CSC represents non-stem cancer cell phenotypic state. Even though this two-phenotypic assumption simplifies the biological complexity of highly diverse phenotypes in cancer, the two-phenotypic setting has been proved as an effective and reasonable simplification for highlighting the minimal process of phenotypic plasticity \cite{dos2013possible,dos2013noise,wang2014dynamics,leder2014mathematical}.
Similar bidirectional transition cascade models were also studied in bacterial community
\cite{Pei2015Fluctuation,Mao2015Slow}.

We now present the cellular process of the two-phenotypic model. From probabilistic point of view, this model can be seen as a discrete-time two-type branching process
\cite{haccou2005branching}. Each cell lives for a fixed time (suppose one unit of time). At the moment of death it gives birth to two daughter cells. More specifically, for each
CSC, it gives birth to two identical CSC daughter cells with probability $\alpha$ (symmetric division), otherwise (with probability $1-\alpha$) it gives birth to one CSC daughter cell and one non-CSC daughter cell (asymmetric division). For each non-CSC, it divides symmetrically into two non-CSC daughter cells with probability $1-\beta$, whereas it divides asymmetrically
into one non-CSC daughter cell and one CSC daughter cell with probability $\beta$ (de-differentiation).
The model will reduce to conventional hierarchical model if letting $\beta=0$, i.e. de-differentiation is not allowed to happen. Hence the model selection with respect to $\beta$ provides an efficient way to evaluate the significance of phenotypic plasticity.

The statistical inference of branching processes has been studied for a long time
\cite{Guttorp1991Statistical}. The usage of statistical methods strongly depends on the data types available. Normally, the observation of the whole genealogy tree generated from underlying
process is quite difficult to obtain (except in very limited experiments
\cite{Hu2015Bayesian}). Instead, only can the numbers or relative frequencies of distinct cell types
be recorded at given moment, and comparatively it is easier to collect
data on relative frequencies than absolute numbers of given cell types \cite{yakovlev2009relative}.
Thus developing statistical approaches for proportion data has a wider range of application.
In this work our proposed method is used for the time-series data on relative frequencies of CSC phenotypic state.

Let $x_A(t)$ be the frequency of CSC state at time $t$,  $\mu_t$ be the expectation of $x_A(t)$, i.e. $\mu_t=\textbf{E}(x_A(t))$, and $\sigma^2_t$ be the variance of $x_A(t)$, i.e. $\sigma^2_t=\textbf{Var}(x_A(t))$. Then we can obtain two important recurrence formulas as follows
  (see \ref{appendixA} for more details):
  \begin{equation}
  \mu_{t+1}=\frac{1+\alpha-\beta}{2}\mu_t+\frac{\beta}{2},
  \label{CE}
  \end{equation}
  \begin{equation}
  \sigma^2_{t+1}=(\frac{1+\alpha-\beta}{2})^2\sigma^2_t+\frac{1}{N_0\times 2^{t+2}}\left\{ [\alpha(1-\alpha)-\beta(1-\beta)] \mu_t + \beta(1-\beta)\right\},
  \label{CV}
  \end{equation}
where $N_0$ is the initial number of the whole population. In next section, we will put forward a Bayesian statistical framework based on the above two formulas.

\subsection{Bayesian framework}\label{bayes}

The data we deal with looks like $\{(m_0, v_0), (m_1, v_1), \cdots, (m_t,v_t),\cdots(m_T, v_T)\}$, where
$m_t$ and $v_t$ are sample mean and sample variance of $n$ realizations of $x_A(t)$ respectively. The Markov property of the model implies that the distribution of data at time $t$ only depends on data at time $t-1$. Thus, we can write the joint likelihood of data as follows:
\[L(\alpha,\beta|m_0,v_0,m_1,v_1,\cdots,m_{T},v_{T})=f(m_0,v_0)\times
f(m_1,v_1|m_0,v_0,\alpha,\beta)\times\cdots\times f(m_{T},v_{T}|m_{T-1},v_{T-1},\alpha,\beta).\]
Note that $f(m_0,v_0)$ is irrelevant to the parameters $\alpha$ and $\beta$, the joint likelihood
can be expressed as
\[L(\alpha,\beta|m_0,v_0,m_1,v_1,\cdots,m_{T},v_{T})\propto
f(m_1,v_1|m_0,v_0,\alpha,\beta)\times\cdots\times f(m_{T},v_{T}|m_{T-1},v_{T-1},\alpha,\beta).\]
According to the asymptotic normality of $x_A(t)$ provided $N_0\gg 1$ (see Theorem 1 in \cite{yakovlev2009relative}),
the asymptotic distributions of $m_t$ and $v_t$ can be given as:\\
\[m_{t}\sim N(\mu_{t},\frac{\sigma^2_{t}}{n}),\]
\[v_{t}\sim\frac{\sigma^2_{t}}{n-1}\chi^2(n-1).\]
Since $m_{t},v_{t}$ are unbiased estimators of $\mu_{t},\sigma^2_{t}$ respectively, we can substitute $\mu_{t}$ and $\sigma^2_{t}$ with the corresponding observed sample mean $m_{t}$ and sample variance $v_{t}$ in the recurrence formulas \eqref{CE} and \eqref{CV}. Given the priors of $\alpha,\beta$ as $Beta(1,1)$, i.e. uniform distribution, we can obtain the following posterior distribution:
\begin{eqnarray*}
&&p(\alpha,\beta|m_0,v_0,m_1,v_1\cdots,m_{T},v_{T})=L(\alpha,\beta|m_0,v_0,m_1,v_1,\cdots,m_{T},v_{T})\times p(\alpha)\times p(\beta)\\
&\propto&\prod_{t=0}^{T-1}\left\{\left[\left(\frac{1+\alpha-\beta}{2}\right)^2v_{t}+\frac{1}{N_0\times 2^{t+2}}\left\{[\alpha(1-\alpha)-\beta(1-\beta)]m_{t}+\beta(1-\beta)\right\}\right]^{-\frac{n}{2}}v_{t+1}^{\frac{n-3}{2}}\right\}\times\\
&&\exp\left\{-\frac{1}{2}\sum_{t=0}^{T-1}\frac{n(m_{t+1}-\frac{1+\alpha-\beta}{2}m_{t}-\frac{\beta}{2})^2}{(\frac{1+\alpha-\beta}{2})^2v_{t}+\frac{1}{N_0\times 2^{t+2}}\left\{ [\alpha(1-\alpha)-\beta(1-\beta)]m_{t}+\beta(1-\beta)\right\}}+\right.\\
&&\left.\frac{(n-1)v_{t+1}}{(\frac{1+\alpha-\beta}{2})^2v_{t}+\frac{1}{N_0\times 2^{t+2}}\left\{[\alpha(1-\alpha)-\beta(1-\beta)]m_{t}+\beta(1-\beta)\right\}}\right\}
\end{eqnarray*}
Note that $N_0\gg1$, the above posterior distribution can be revised as the following simplified expression:
\begin{eqnarray*}
p(\alpha,\beta|m_0,v_0,m_1,v_1\cdots,m_{T},v_{T})&\propto&\prod_{t=0}^{T-1}\left\{\left[\left(\frac{1+\alpha-\beta}{2}\right)^2v_{t}\right]^{-\frac{n}{2}}v_{t+1}^{\frac{n-3}{2}}\right\}\times\\
&&\exp\left\{-\frac{1}{2}\sum_{t=0}^{T-1}\frac{n(m_{t+1}-\frac{1+\alpha-\beta}{2}m_{t}-\frac{\beta}{2})^2}{(\frac{1+\alpha-\beta}{2})^2v_{t}}+\frac{(n-1)v_{t+1}}{(\frac{1+\alpha-\beta}{2})^2v_{t}}\right\}
\end{eqnarray*}
We will show in Section \ref{simu2} that $N_0$ does has little impact on the simulation results, and then validate the above simplified posterior distribution.

Based on the proposed posterior distribution, $\alpha$ and $\beta$ are sampled iteratively by Gibbs sampling and MH algorithm. We draw several sample chains independently and apply MPSRF \cite{gelman1992inference,Brooks1998general} to check the convergence of MCMC chains. If the MCMC chains converge, we obtain the point estimation and interval estimation as mean value and interval between $2.5\%$ and $97.5\%$ quantiles of converged posterior samples respectively.

\section{Results}\label{result}

In this section we perform some simulations to validate our algorithm and also apply our method to a published data set of SW620 colon cancer cell line \cite{yang2012dynamic}.

\subsection{Simulation I}\label{simu1}

In the first simulation, we generate synthesized data sets to test the performance of our algorithm. In particular, we set $N_0=10^4$ and generate parameters $\alpha$ and $\beta$ from their priors respectively, drawing mean value $m_0$ from $Unif(0.3,0.7)$ and standard deviance $v_0^{\frac{1}{2}}$ from $Unif(0.01,0.03)$. We then synthesize $\{m_1,v_1\},\cdots,\{m_{T},v_{T}\}$ sequentially based on the conditional distribution function in Section~\ref{bayes}. In all we generate 100 groups of parameters and each group consists of 20 simulations. Table~\ref{res1} demonstrates the estimation results including Mean Square Error (MSE) of point estimation, averaged length of $95\%$ confidence intervals (AL) and mean proportion of interval estimation covering the true value of parameters (CR).
\begin{table}[!h]
\centering
\begin{tabular}{|c|c|c|c|c|c|}
\hline
\multicolumn{3}{|c|}{$\alpha$}&\multicolumn{3}{c|}{$\beta$}\\
\cline{1-6}
MSE&AL&CR&MSE&AL&CR\\
\hline
0.0004&0.0414&0.924&0.0003&0.0321&0.917\\
\hline
\end{tabular}
\caption{Estimation results of Simulation I.}\label{res1}
\end{table}
It is easy to find that our algorithm is accurate for the synthesized data sets, with small MSE, high coverage rate and narrow confidence interval.

\subsection{Simulation II}\label{simu2}

In the second simulation, we generate data sets by following the cellular processes of the two-phenotypic model described in Section~\ref{model} to validate our method.
We set $N_0=10^4$ and randomly generate the true values of parameters $\alpha$ and $\beta$ from their priors. Given each value of $(\alpha,\beta)$ and
initial state $x_A(0)$ sampled from a Normal distribution $N(\mu',\sigma'^2)$ with $\mu'\sim Unif(0.3,0.7)$ and $\sigma'\sim Unif(0.01,0.03)$,
we generate $n=5$ synthesized realizations of $x_A(t)$ and then calculate the sample mean $m_t$ and sample variance $v_t$ as algorithm input.
In all we generate 100 groups of parameters $(\alpha,\beta)$ with each consisting of 20 simulations. Table~\ref{res2} demonstrates the MSE, AL and CR of
two methods with and without $N_0$ as known true value.
\begin{table}[!h]
\centering
\begin{tabular}{|c|c|c|c|c|c|c|}
\hline
\multirow{2}{*}{Set}&\multicolumn{3}{|c|}{$\alpha$}&\multicolumn{3}{c|}{$\beta$}\\
\cline{2-7}
&MSE&AL&CR&MSE&AL&CR\\
\hline
With $N_0$&0.0004&0.0380&0.979&0.0002&0.0302&0.979\\
\hline
Without $N_0$&0.0006&0.0378&0.962&0.0003&0.0296&0.963\\
\hline
\end{tabular}
\caption{Estimation results of Simulation II.}\label{res2}
\end{table}
From the table, we can see that our algorithm well captures the cellular process when only mean and variance data are available. Both point estimation and interval estimation achieve high accuracy even if $N_0$ is unknown in our algorithm. An example of estimation results can be found in \ref{estimation}.

\subsection{Real data}

Our real data application is based on a published data of SW620 colon cancer cell line
(see Figure 4A in \cite{yang2012dynamic}). There are four groups of data in this
experiment. In each group, CSC proportions were measured via FACS (Fluorescence-activated cell sorting),
and both sample mean $m_t$ and sample variance $v_t$ were recorded successively.
The four groups differ in the initial states of relative frequencies:
(A) $0.6\%$ CSCs and $99.4\%$  non-CSCs;
(B) 70\% CSCs and 30\% non-CSCs;
(C) $99.4\%$ CSCs and $0.6\%$ non-CSCs;
(D) $65.4\%$ CSCs and $34.6\%$ non-CSCs.
We assume four sets of parameters $\Theta_1=\{\alpha_1,\beta_1\},\Theta_2=\{\alpha_2,\beta_2\},\Theta_3=\{\alpha_3,\beta_3\},\Theta_4=\{\alpha_4,\beta_4\}$ corresponding to four groups of data and perform model selection by calculating Deviance information criterion (DIC, \cite{Gelman2003Bayesian}).
As a generalization of Akaike information criterion (AIC) and Bayesian information criterion (BIC), DIC is also an estimator of the relative quality of statistical models for given data, and it is particularly useful in Bayesian statistical settings, representing the trade-off between the error of fitting and the complexity of the model.
The smaller the DIC is, the more favorable the model is.

Our primary concern is whether the hypothesis of reversible phenotypic plasticity is
superior to the hypothesis of cellular hierarchy given the colon cancer data.
Note that the phenotypic plasticity model will reduce to the hierarchical model
by setting $\beta_i=0$ ($i=1, 2, 3, 4$).
We would like to compare the DIC values between the full model and the hierarchical model without de-differentiation.
Our result shows that the DIC value of the full model is $-103.23$, which is smaller than that of the hierarchical model
without de-differentiation ($-84.05$). Model selection thus suggests that de-differentiation significantly
improve the quality of the model for the given data.

Furthermore, we are interested in whether some of the four groups of data share the common values of the parameters, from which
we can see the dependency of the parameters on the initial states.
The DIC values of different models are shown in Table~\ref{res3} (see \ref{para} for the estimated values of the parameters). From the table, we can find that models $\Theta_1\ne\Theta_2\ne\Theta_3\ne\Theta_4$ and $\Theta_1=\Theta_2$ obtain the smallest DIC values, which means these two models are favorable. Due to the limited data length, these two models cannot be distinguished since the difference between their DIC values is only about 0.15, without statistical significance. It is interesting that the model $\Theta_1\ne\Theta_2\ne\Theta_3\ne\Theta_4$ is selected as the favorable model. That is, none of the four groups share the common parameters, all the parameters are sensitive to the initial states of relative frequencies. Note that different initial relative frequencies correspond to different states of phenotypic heterogeneity at the beginning of cell sorting, our result
suggests an interesting heterogeneity-dependency of the phenotypic plasticity model.

\begin{table}[!h]
\centering
\begin{tabular}{|c|c|}
\hline
Model&DIC\\
\hline
$\Theta_1\ne\Theta_2\ne\Theta_3\ne\Theta_4$&-103.23\\
\hline
$\Theta_1=\Theta_2$&-103.38\\
\hline
$\Theta_1=\Theta_3$&7.03\\
\hline
$\Theta_1=\Theta_4$&-90.00\\
\hline
$\Theta_2=\Theta_3$&125.37\\
\hline
$\Theta_2=\Theta_4$&-82.37\\
\hline
$\Theta_3=\Theta_4$&71.72\\
\hline
$\Theta_1=\Theta_2,\Theta_3=\Theta_4$&71.57\\
\hline
$\Theta_1=\Theta_3,\Theta_2=\Theta_4$&27.89\\
\hline
$\Theta_1=\Theta_4,\Theta_2=\Theta_3$&138.60\\
\hline
$\Theta_2=\Theta_3=\Theta_4$&173.39\\
\hline
$\Theta_1=\Theta_3=\Theta_4$&96.39\\
\hline
$\Theta_1=\Theta_2=\Theta_4$&-83.58\\
\hline
$\Theta_1=\Theta_2=\Theta_3$&146.96\\
\hline
$\Theta_1=\Theta_2=\Theta_3=\Theta_4$&189.42\\
\hline
\end{tabular}
\caption{DIC of Bayesian model selection.}\label{res3}
\end{table}

\section{Conclusions}
\label{conclusion}

We have presented a Bayesian statistical analysis on a stochastic phenotypic plasticity model of cancer cells.
Both simulation studies and real data analysis have shown the power of our method. Compared to the deterministic models
in previous studies \cite{zhou2014multi,wang2014dynamics}, the stochastic model here equipping
with statistical inference makes quantitative modeling in a more sophisticated way.
On one hand, parameter estimation is extended from point-level into interval-level.
On the other hand, model selection provides a systematic means to compare different
models, and then helps to evaluate different biological hypotheses.

It should be noted that, the two-phenotypic model is a very simplified model.
By focusing on our attention to the reversibility between CSCs and non-CSCs, many
biologically complex mechanisms are not included in this model. For example, the model
is discrete-time, ignoring the complicated time distribution of cell cycle
\cite{Mode1971Multitype,Golubev2010Exponentially,Jiang2017Phenotypic}.
Developing feasible Bayesian framework for more complicated models
could be of great value in future researches.

\section*{Acknowledgements}

This work is supported by the National Natural Science
Foundation of China (Grant Nos. 11601453 and 11401499), the Natural
Science Foundation of Fujian Province of China (Grant Nos.
2017J05013 and 2015J05016), the Fundamental Research Funds for the
Central Universities in China (Grant No. 20720160004).

\appendix

\section{Derivation of Eqs. \eqref{CE} and \eqref{CV}}
\label{appendixA}

Here we present the mathematical details of how to obtain Eqs. \eqref{CE} and \eqref{CV}, i.e. the recurrence formulas of the expectation and variance
of $x_A(t)$.

Let $n_A(t)$ and $n_B(t)$ are the numbers of CSC and non-CSC states, $x_A(t)$ and $x_B(t)$ are the frequencies of CSC and non-CSC states,
$N(t)$ is the population size of the whole population. It is easy to know that $x_A(t)=1-x_B(t)$, $x_A(t)=n_A(t)/N(t)=n_A(t)/(n_A(t)+n_B(t))$, and
$x_B(t)=n_B(t)/N(t)$. Suppose

$$
\xi_i=\left\{
      \begin{array}{ll}
        1, & \hbox{with probability $\alpha$} \\
        0, & \hbox{with probability $1-\alpha$}
      \end{array}
    \right.
$$

$$
\eta_i=\left\{
      \begin{array}{ll}
        1, & \hbox{with probability $\beta$} \\
        0, & \hbox{with probability $1-\beta$,}
      \end{array}
    \right.
$$
then we have
$$n_A(t+1)=n_A(t)+\sum_{i=1}^{n_A(t)}\xi_i+\sum_{i=1}^{n_B(t)}\eta_i.$$
By taking conditional expectation on both sides we have
$$E(n_A(t+1)|n_A(t))=n_A(t)+n_A(t)\times\alpha+n_B(t)\times\beta.$$
Then for the conditional expectation of $x_A(t)$ we have
\begin{eqnarray*}
E(x_A(t+1)|x_A(t))&=&E(\frac{n_A(t+1)}{N(t+1)}|x_A(t))\\
&=&E(\frac{n_A(t+1)}{2N(t)}|x_A(t))\\
&=&\frac{1}{2}\frac{1}{N(t)}(n_A(t)+n_A(t)\times\alpha+n_B(t)\times\beta)\\
&=&\frac{1}{2}(x_A(t)+x_A(t)\times\alpha+x_B(t)\times\beta)\\
&=&\frac{1}{2}(1+\alpha) x_A(t)+\frac{1}{2}(1-x_A(t))\beta\\
&=&\frac{1+\alpha-\beta}{2}x_A(t)+\frac{\beta}{2}.\\
\end{eqnarray*}
With the law of total expectation we obtain Eq. \eqref{CE} as follows
\begin{equation*}
\mu_{t+1}=E(x_A(t))=E(E(x_A(t+1)|x_A(t)))=\frac{1+\alpha-\beta}{2}\mu_t+\frac{\beta}{2}.
\end{equation*}
For the conditional variance of $x_A(t)$,
\begin{eqnarray*}
Var(x_A(t+1)|x_A(t))&=&Var(\frac{n_A(t+1)}{N(t+1)}|x_A(t))\\
&=&Var(\frac{n_A(t)+\sum_{i=1}^{n_A(t)}\xi_i+\sum_{i=1}^{n_B(t)}\eta_i}{N(t+1)}|x_A(t))\\
&=&Var(\frac{\sum_{i=1}^{n_A(t)}\xi_i+\sum_{i=1}^{n_B(t)}\eta_i}{N_0\times2^{t+1}}|x_A(t))\\
&=&\frac{1}{N_0^2\times2^{2t+2}}(n_A(t)\alpha(1-\alpha)+n_B(t)\beta(1-\beta))\\
&=&\frac{1}{N_0\times2^{t+2}}(\frac{n_A(t)}{N_0\times2^t}\alpha(1-\alpha)+\frac{n_B(t)}{N_0\times2^t}\beta(1-\beta))\\
&=&\frac{1}{N_0\times2^{t+2}}(x_A(t)\alpha(1-\alpha)+x_B(t)\beta(1-\beta))
\end{eqnarray*}
With the law of total variance we obtain Eq. \eqref{CV} as follows
\begin{eqnarray*}
\sigma^2_{t+1}&=&Var(E(x_A(t+1)|x_A(t)))+E(Var(x_A(t+1)|x_A(t)))\\
&=&(\frac{1+\alpha-\beta}{2})^2\sigma^2_t+\frac{1}{N_0\times 2^{t+2}}\left\{ [\alpha(1-\alpha)-\beta(1-\beta)] \mu_t + \beta(1-\beta)\right\}.
\end{eqnarray*}

\section{An example of simulation results}\label{estimation}
\begin{figure}
\begin{minipage}[t]{0.5\linewidth}
\centering
\includegraphics[width=3in]{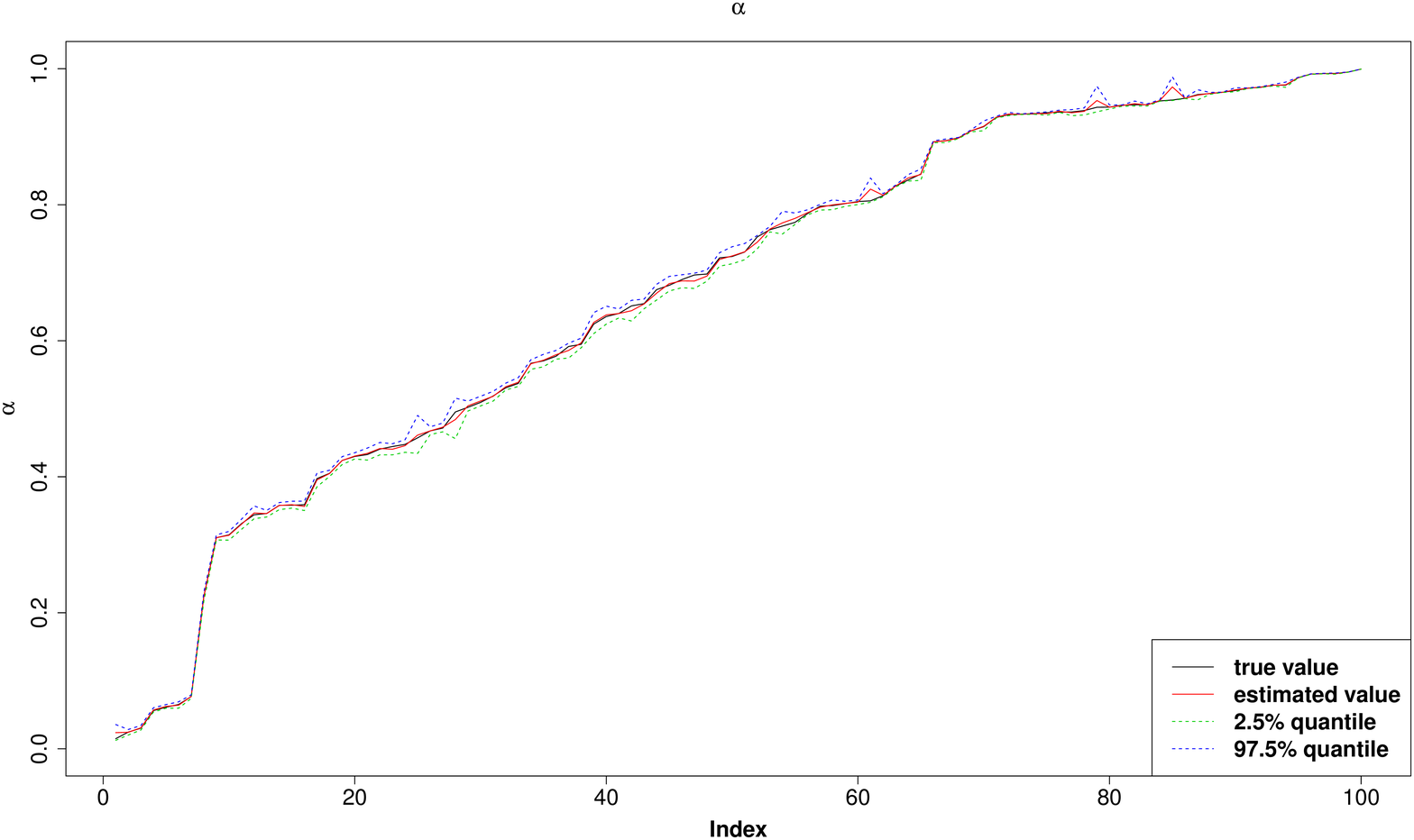}
\centerline{(a) Simulation I: $\alpha$}
\end{minipage}
\begin{minipage}[t]{0.5\linewidth}
\centering
\includegraphics[width=3in]{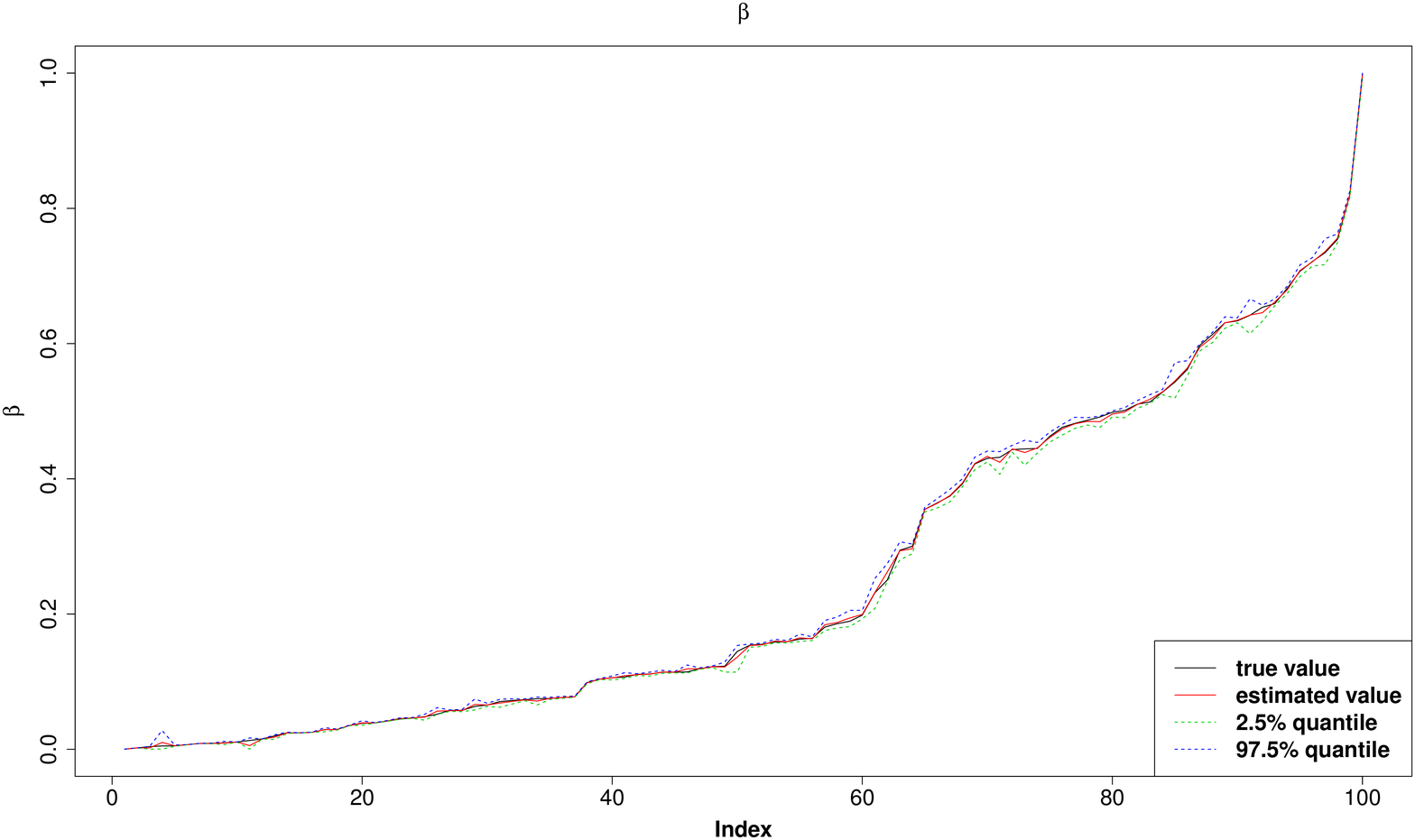}
\centerline{(b) Simulation I: $\beta$}
\end{minipage}
\begin{minipage}[t]{0.5\linewidth}
\centering
\includegraphics[width=3in]{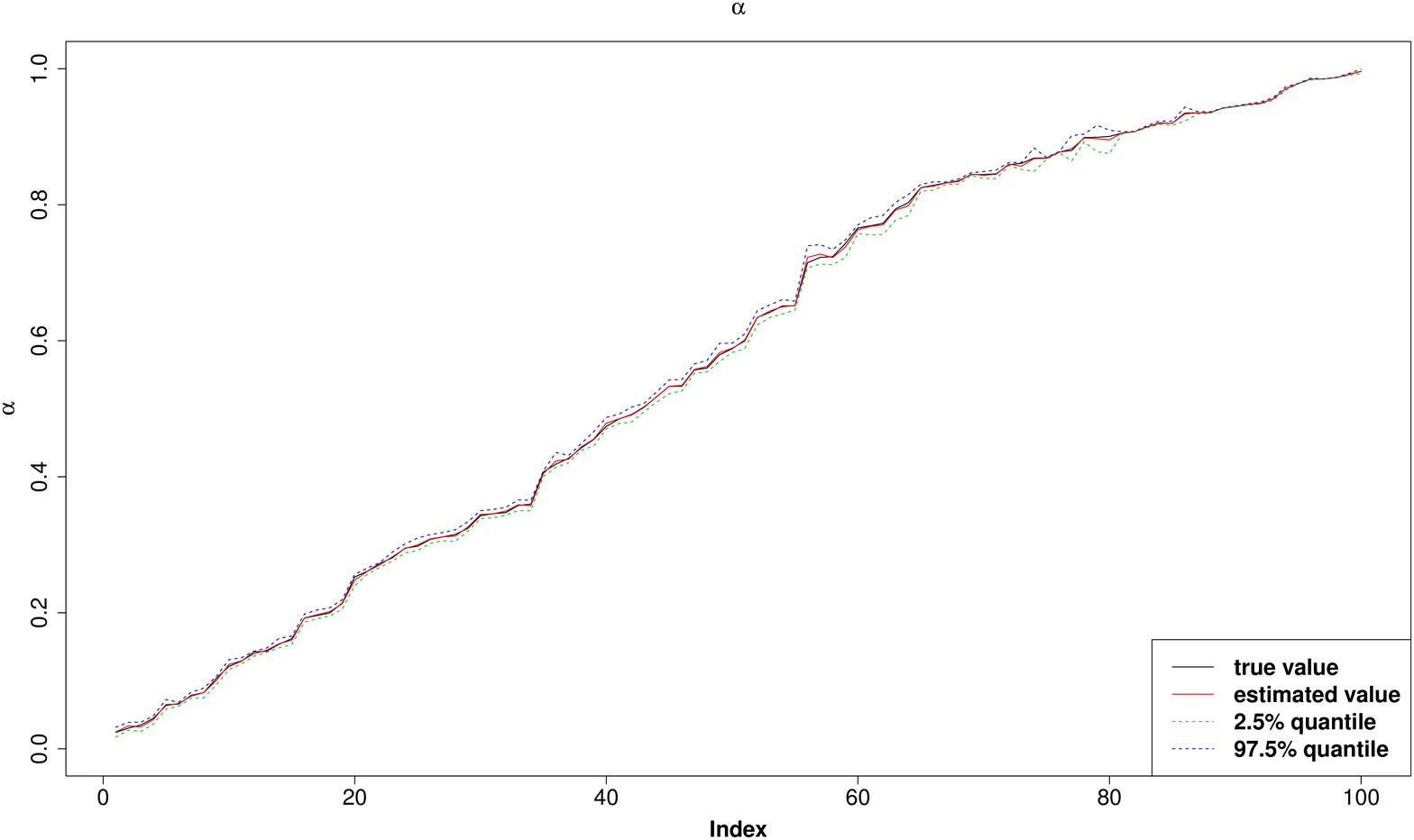}
\centerline{(c) Simulation II with $N_0$: $\alpha$}
\end{minipage}
\begin{minipage}[t]{0.5\linewidth}
\centering
\includegraphics[width=3in]{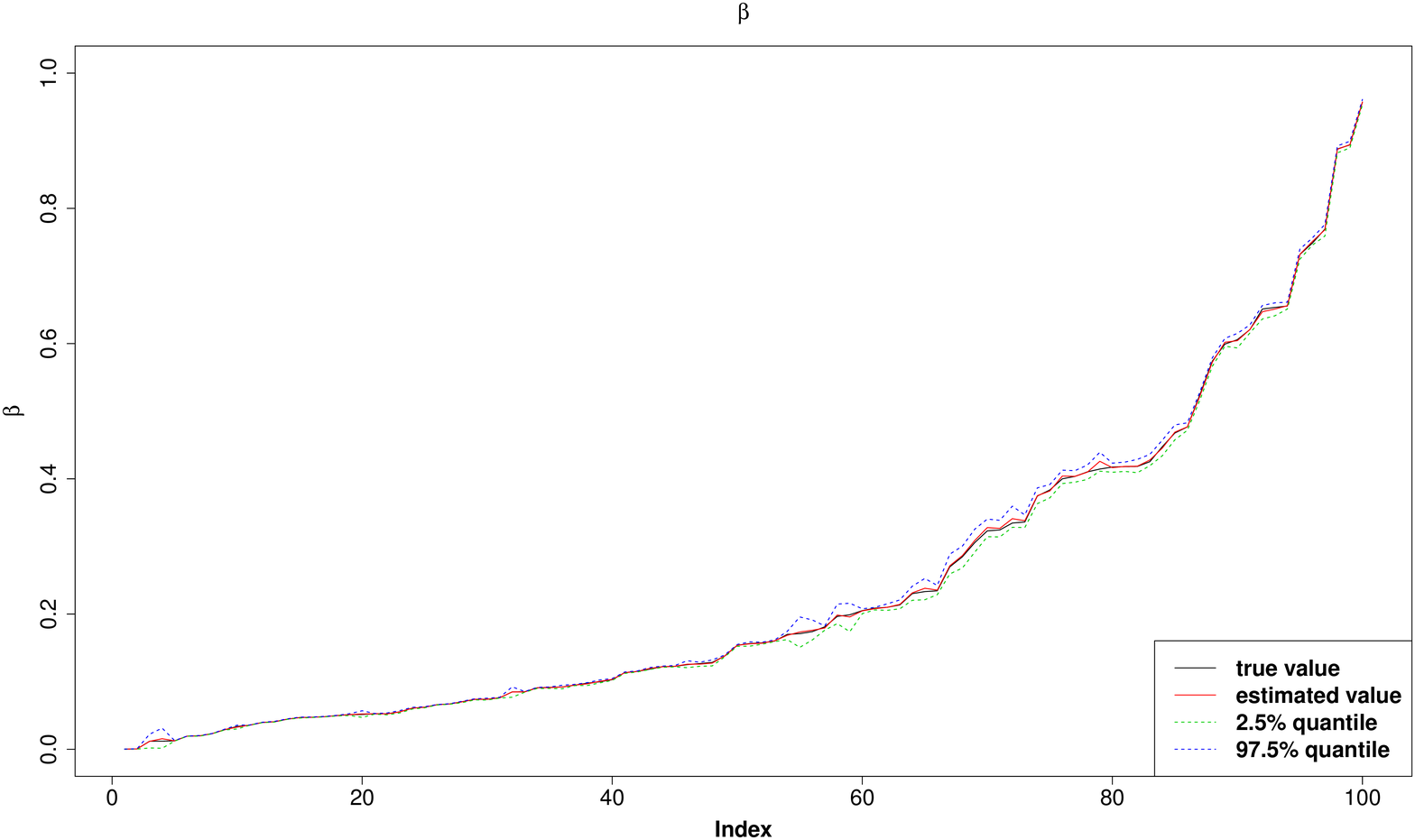}
\centerline{(d) Simulation II with $N_0$: $\beta$}
\end{minipage}
\begin{minipage}[t]{0.5\linewidth}
\centering
\includegraphics[width=3in]{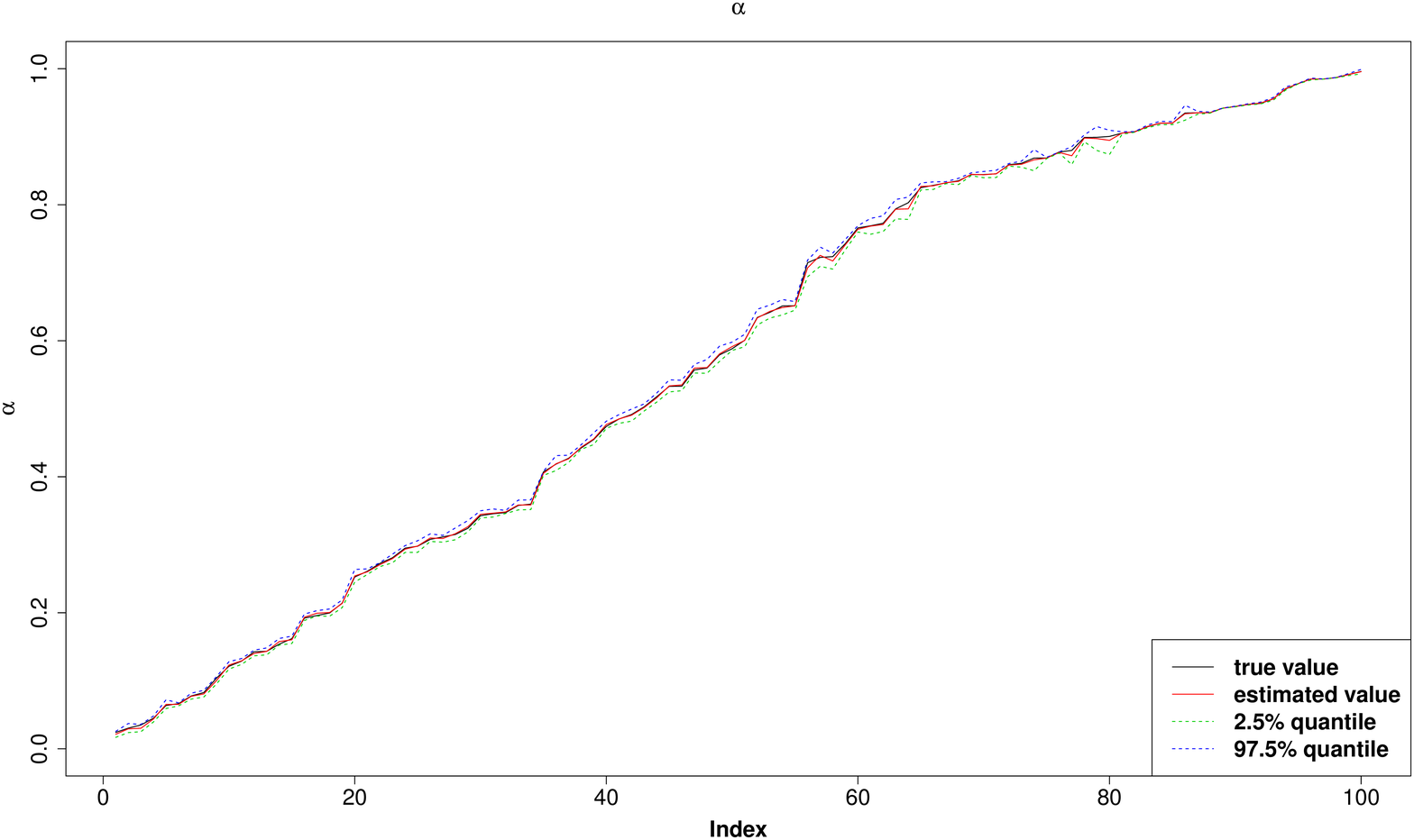}
\centerline{(e) Simulation II without $N_0$: $\alpha$}
\end{minipage}
\begin{minipage}[t]{0.5\linewidth}
\centering
\includegraphics[width=3in]{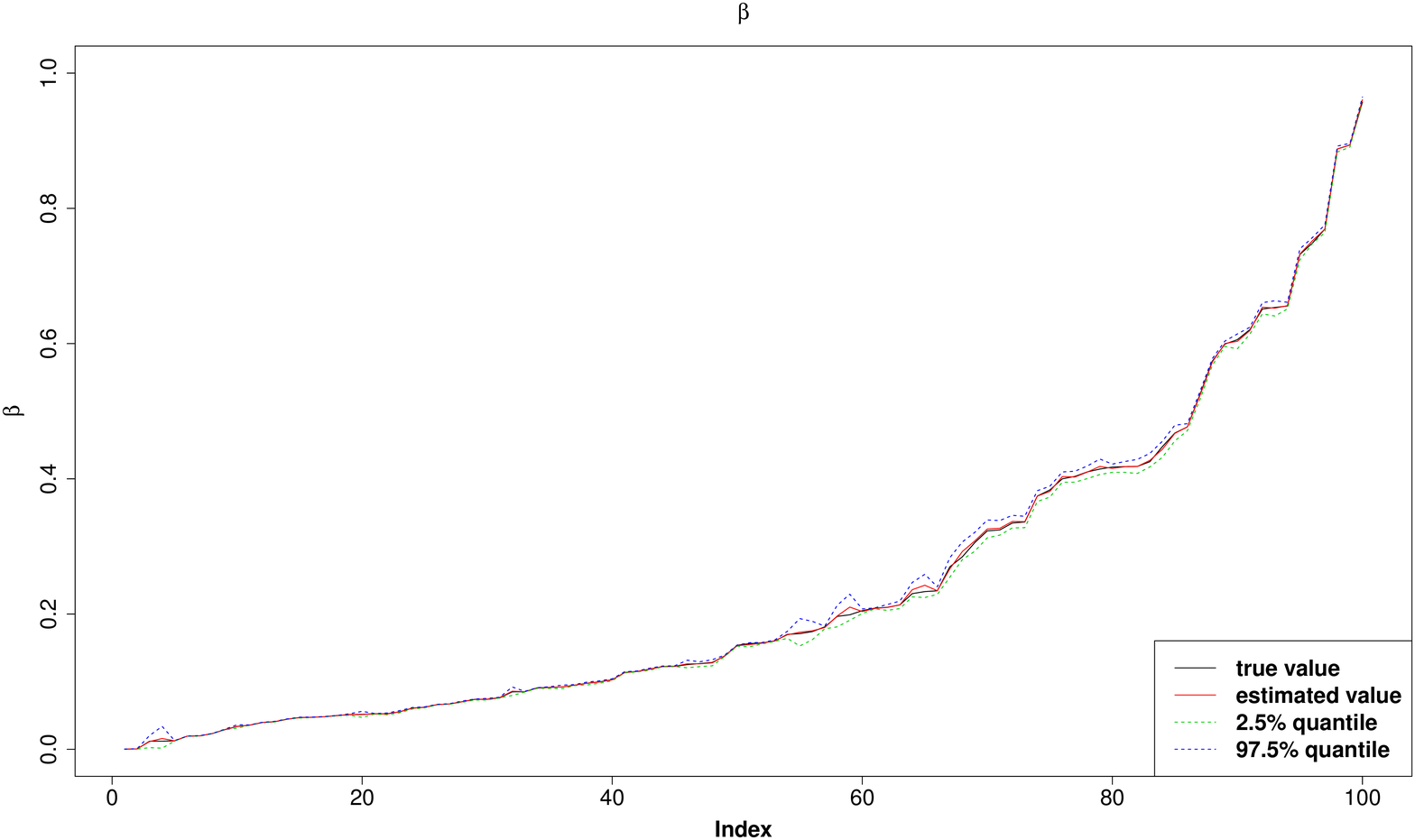}
\centerline{(f) Simulation II without $N_0$: $\beta$}
\end{minipage}
\caption{An example of estimation results in Simulation I and Simulation II. Black solid lines represent true values of parameters and red lines represent point estimation. And two dashed lines demonstrate interval estimations. For Simulation II, both the estimation results with and without $N_0$ are shown}\label{s1}
\end{figure}

We select one simulation from each of the 20 parameters settings in Simulations I$\&$II, and show an example of estimation results in Figure~\ref{s1}.

\section{Point estimation of real data}\label{para}
Here we show point estimation results of real data for different models.

\begin{table}[!h]
\centering
\begin{tabular}{|c|c|c|}
\hline
Model&$\alpha$&$\beta$\\
\hline
$\Theta_1$&0.978&0.125\\
\hline
$\Theta_2$&0.990&0.118\\
\hline
$\Theta_3$&0.890&0.008\\
\hline
$\Theta_4$&0.991&0.014\\
\hline
$\Theta_1=\Theta_2$&0.987&0.121\\
\hline
$\Theta_1=\Theta_3$&0.889&0.056\\
\hline
$\Theta_1=\Theta_4$&0.992&0.031\\
\hline
$\Theta_2=\Theta_3$&0.899&0.072\\
\hline
$\Theta_2=\Theta_4$&0.993&0.036\\
\hline
$\Theta_3=\Theta_4$&0.902&0.064\\
\hline
$\Theta_1=\Theta_2=\Theta_3$&0.882&0.154\\
\hline
$\Theta_1=\Theta_2=\Theta_4$&0.996&0.042\\
\hline
$\Theta_1=\Theta_3=\Theta_4$&0.888&0.131\\
\hline
$\Theta_2=\Theta_3=\Theta_4$&0.900&0.116\\
\hline
$\Theta_1=\Theta_2=\Theta_3=\Theta_4$&0.892&0.152\\
\hline
\end{tabular}
\caption{Parameter estimation of different models. $\Theta_i$ represents the estimation of parameters of $i$-th group in real data and $\Theta_{i_1}=\cdots=\Theta_{i_k}$ represents estimation of common parameters within these $k$ groups.}\label{}
\end{table}





\bibliographystyle{elsarticle-num}






\end{document}